\documentclass[]{llncs}
\usepackage[T1]{fontenc}
\usepackage{graphicx}
\usepackage{algorithmic}
\usepackage{amsmath}
\usepackage{multirow}
\usepackage{marvosym}

\begin{document}
%
% --- 标题部分 ---
\title{ML-SAN: Multi-Level Speaker-Adaptive Network for Emotion Recognition in Conversations}

% --- 作者部分 ---
%\author{Kexue Wang \and
%Yinfeng Yu \and
%Liejun Wang}

\author{
	Kexue Wang$^{1,2,3}$, Yinfeng Yu$^{1,2,3}$, and Liejun Wang$^{1,2,3,}$%
	\thanks{Liejun Wang is the corresponding author. (E-mail: wljxju@xju.edu.cn). This research was financially supported by the National Natural Science Foundation of China (Grant numbers 62472368, and 62463029).}%
}

% 运行头（页眉显示的简写作者名）
\authorrunning{K. Wang et al.}

% --- 单位与邮箱部分 ---
\institute{\small $^1$Joint Research Laboratory for Embodied Intelligence, Xinjiang University\\
$^2$Joint International Research Laboratory of Silk Road Multilingual Cognitive Computing, Xinjiang University\\
$^3$School of Computer Science and Technology, Xinjiang University, Urumqi 830017, China}

\maketitle

\begin{abstract}
To establish empathy with machines, it is essential to fully understand human emotional changes. However, research in multimodal emotion recognition often overlooks one problem: individual expressive traits vary significantly, which means that different people may express emotions differently. In our daily lives, we can see this. When communicating with different people, some express "happiness" through their facial expressions and words, while others may hide their happiness or express it through their actions. Both are expressions of 'happiness,' but such differences in emotional expression are still too difficult for machines to distinguish. Current emotion recognition remains at a 'static' level, using a single recognition model to identify all emotional styles. This "simplification" often affects the recognition results, especially in multi-turn dialogues. To address this problem, this paper introduces a novel Multi-Level Speaker Adaptive Network (ML-SAN), which, specifically, effectively addresses the challenge of speaker identity information confusion. ML-SAN does not simply assign a speaker's ID after recognition; instead, it employs a three-stage adaptive process: First, Input-level Calibration uses Feature-Level Linear Modulation (FiLM) to adjust the raw audio and visual features into a neutral space unrelated to the speaker. Then, Interaction-level Gating re-adjusts the trust level for each modality (e.g., voice or facial features) based on the speaker's identity information. Finally, Output-level Regularization maintains the consistency of speaker features in the latent space. Tests on the MELD and IEMOCAP datasets show that our model (ML-SAN) achieves better results, performs exceptionally well in handling challenging tail sentiment categories, and better addresses the diversity of speakers in real-world scenarios.

\end{abstract}
\section{Introduction}
\begin{figure} % [t] 表示允许浮动到本页或下一页的顶部
    \centering
    % 将宽度设为 0.85\columnwidth (85%的栏宽)，比之前小一圈
    \includegraphics[width=0.95\columnwidth]{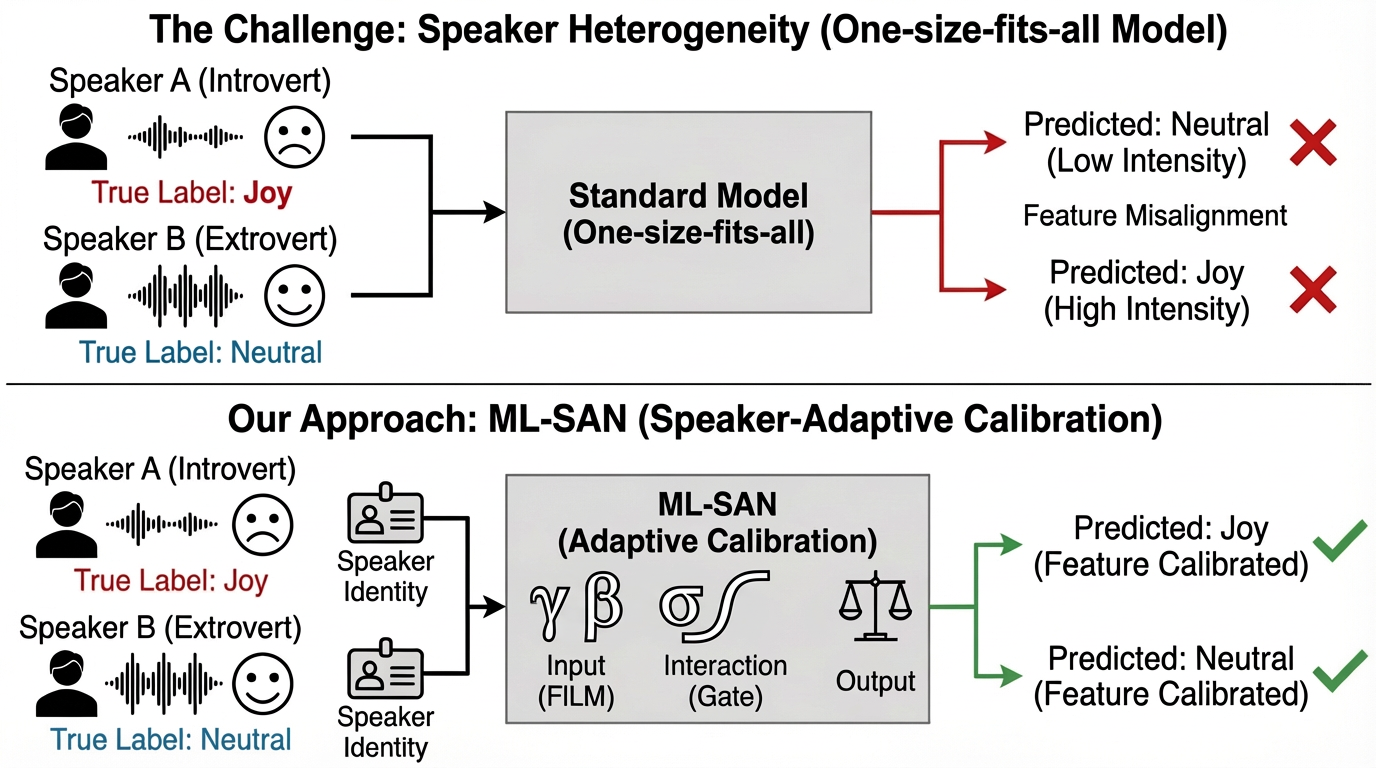} 
    
    \caption{The challenge of speaker heterogeneity and our solution.}
    \label{fig:A}
\end{figure}

“The question is not whether intelligent machines
 can have emotions, but whether machines without
 emotions can achieve intelligence”, as mentioned
 in “Society of Mind” \cite{minsky1985society} . Empowering
 machines with the ability to understand emotions in
 various scenarios has always been the unwavering
 direction of researchers.
 In contrast to conventional binary sentiment analysis tasks \cite{su2025multimodal} , which only rely on text with explicit attitude tendencies, the emotion recognition in conversation task aims
 to identify more fine-grained emotional tendencies
 in each sentence of a conversation. Specifically, for
 a given complete dialogue sequence input and a set
 of emotional labels, the model is required to accurately assign an emotional label to each sentence.
 Intuitively, the recognition of emotional tendencies in the target sentence is heavily influenced
 by its historical utterances \cite{wang2025multimodal} ,
 and there is significant variation in how different
 speakers perceive and express emotions \cite{li2025multimodal} . Therefore, it is imperative to meticulously model the speakers and dialogue context. Fig. \ref{fig:A} illustrates this difference: for example, a high-arousal expression from an introvert might be perceived as "neutral" by a global model, whereas an extrovert's "joyful state" might share similar acoustic distributions with another speaker's "anger." 

Most existing methods treat all speakers as interchangeable entities \cite{sun2024selfsupervised} , which leads to two critical failure modes. First, the assumption of speaker interchangeability precipitates \textbf{Feature Misalignment}, where the model fails to establish robust decision boundaries across diverse expressive styles. Consequently, the model struggles to find a common latent space that accurately represents emotions for all individuals. Second, this speaker-agnostic approach results in \textbf{Ineffective Fusion}; that is, the system fails to dynamically prioritize the most informative modalities (e.g., facial cues vs. vocal tone) for a specific speaker, leading to a "fusion failure" where the most expressive features of a particular individual are underutilized.

To rectify this, we introduce the \textbf{M}ulti-\textbf{L}evel \textbf{S}peaker-\textbf{A}daptive \textbf{N}etwork (\textbf{ML-SAN}). Our methodology draws inspiration from conditional computation techniques like Feature-wise Linear Modulation (FiLM) \cite{gu2022multimodal} and multi-task learning paradigms \cite{guo2023multimodal} . Unlike previous work, our model ML-SAN injects speaker identity as an active control signal at three key points:

\begin{itemize}
    \item \textbf{Input Calibration:} First, FiLM is used to dynamically scale the feature distributions obtained from OpenSMILE \cite{yao2021multimodal}  and ResNet \cite{liu2020multimodal} , thereby normalizing speaker bias.
    \item \textbf{Interaction Gating:} We have designed a dynamic gating mechanism that can weight audio and visual inputs based on the speaker's latent features.
    \item \textbf{Output Regularization:} To ensure final results retain the speaker's distinguishing information, we also introduced an auxiliary classification task.
\end{itemize}

\section{Related Works}

\subsection{Context-aware Emotion Recognition}
The core approach to emotion recognition is to model dialogues \cite{zhou2015research} . In early studies on emotion recognition, researchers mainly relied on recurrent neural networks, treating dialogue as nodes on an image. There are also some relatively novel methods in this field, such as LSTM and DialogueGCN \cite{ghosal2019dialoguegcn} , which have presented their own insights and methodologies in emotion recognition \cite{kim2025omer,kulkarni2025emotion} . However, it is undeniable that existing methods still have some issues, such as overlooking the differences between speakers due to varying personalities \cite{kapila2025multimodal} . As discussed earlier, if a model treats different speakers as 'the same,' it will inevitably lead to poor recognition performance.

\subsection{Multimodal Fusion Strategies}
The synergy between acoustic and visual modalities is crucial for robustness. Classic methods such as Emotion recognition based on EEG signals and face images (EEG) \cite{lian2025emotion} and Multimodal fusion: A study on speech-text emotion recognition with the integration of deep learning (MF) \cite{shang2024multimodal} have laid the mathematical foundation for cross-modal interaction. InstructERC \cite{zhao2024instructerc} extend cross-modal interaction to sequential data. With the advent of Transformers, model architectures such as Dual Graph Attention Networks for Emotion Recognition in Conversations(Dual-GATs) \cite{li2023dual} , MMGCN \cite{hu2021mmgcn} ,DialogueRNN \cite{majumder2019dialoguernn} , UniMSE \cite{hu2022unimse} and CDSAGV \cite{poria2017context} have successfully demonstrated the effectiveness of cross-modal attention mechanisms. In the ERC framework, MMGCN, ARL \cite{lin2024review} ,MCER \cite{xue2024multimodal} ,MSERB \cite{fang2024multimodal} and RoMERF \cite{li2024multimodal} , and MultiEMO \cite{li2023multimoemo} have all used these advanced fusion strategies. For other architectures, such as ICON and BERT \cite{gu2022speaker} , interactive fusion is also involved. Recent advances in audio-visual learning have shown promising results, such as audio-visual navigation \cite{YinfengICLR2022saavn,li2025audio,zhang2025advancing,zhang2025iterative,yu2025dynamic} and speech enhancement \cite{mattursun2024bss,cao2024vnet}. Additionally,Missing multimodal sentiment analysis has been studied in \cite{wang2025modality}. However, these strategies are still speaker-independent, \textit{speaker-agnostic}, using the same fusion logic regardless of the speaker's identity.

\subsection{Speaker Modeling in ERC}

In the ERC field, traditional speaker modeling usually only involves simple embedding initialization, such as M3GAT \cite{sun2024m3gat} and COGMEN \cite{joshi2022cogmen} . Although techniques like RMERMM \cite{li2024robust} and AMERABE \cite{wu2024multimodal} have been mentioned in related areas, their adaptability to the ERC field is still far from sufficient \cite{busso2008iemocap,poria2019meld} . 

Unlike these static approaches, ML-SAN shifts the speaker's role from a passive description to an \textbf{active control signal}. By injecting this signal into three stages (calibration, fusion, and regularization), ML-SAN achieves a more granular adaptation to individual expressive styles, effectively addressing the limitations of prior speaker-agnostic or simple-embedding-based frameworks.

\section{Methodology}
In this section, we elaborate on the ML-SAN framework. The main goal is to shift the speaker’s role from passive description to active modulation, thereby guiding subsequent feature processing (see Fig. 2). The core intuition behind ML-SAN is a \textbf{hierarchical adaptation strategy} that addresses speaker heterogeneity at three distinct levels:

\noindent \textbf{(
1) Input Calibration :} This stage acts as a ``de-biasing'' filter
 to resolve \textit{Feature Misalignment}. 
\textbf{(
2) Interaction Gating:} This mechanism serves as
 a dynamic selector to mitigate \textit{Ineffective Fusion}. 
\textbf{(
3) Output Regularization:} This component functions as an identity anchor to prevent information loss.
This multi-stage synergy ensures that the framework interprets emotional cues through the lens of individual speaker characteristics, rather than relying on a generic, speaker-agnostic boundary.

\begin{figure*}
	\centering
	% 将宽度从 0.95 改为 0.8 (或者 0.75)，图片会自动等比例缩小
	\includegraphics[width=0.85\textwidth]{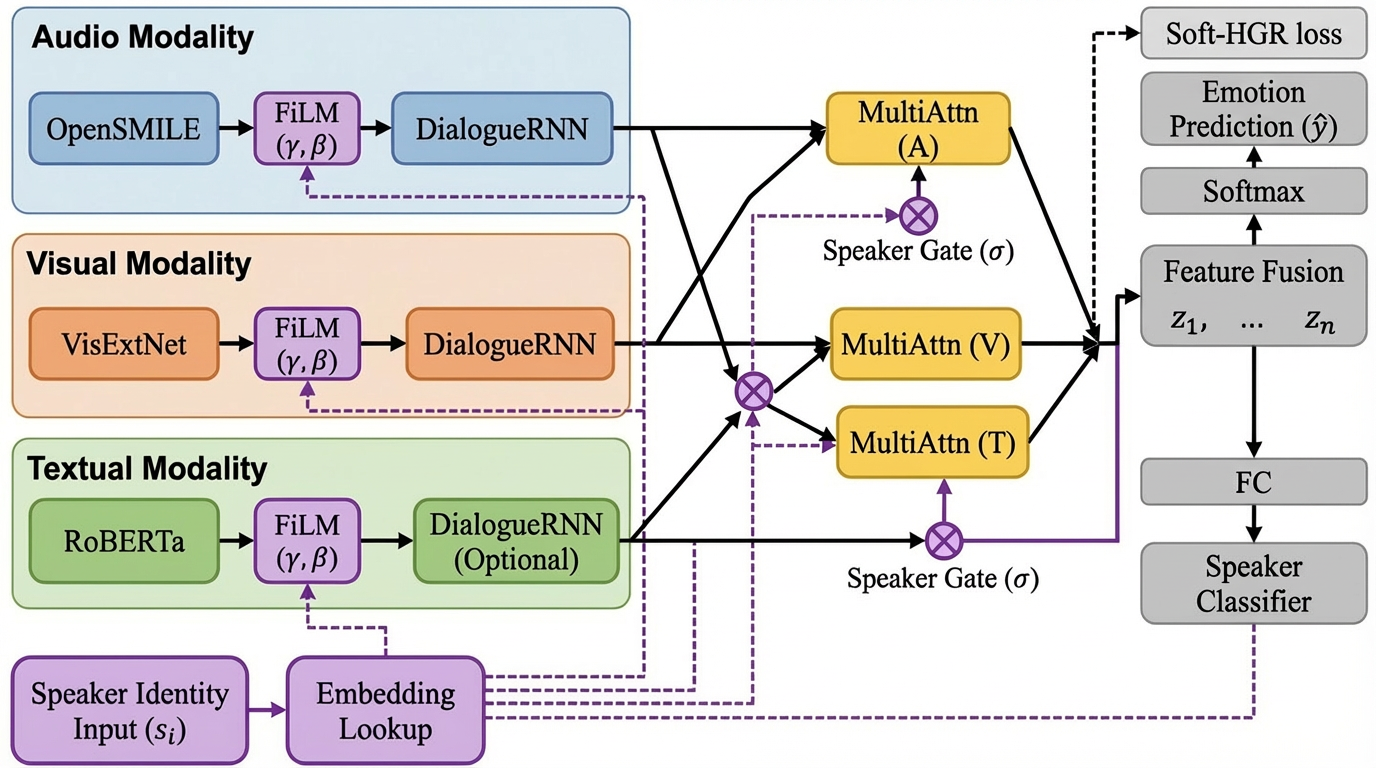}
	
	\caption{The overall architecture of the Multi-Level Speaker-Adaptive Network (ML-SAN).}
	\label{fig:B}
\end{figure*}

\subsection{Input-level: Speaker-Adaptive Feature Modulation}
To address the problem of distribution shift in feature files extracted from pre-trained models due to individual differences, the method we use is FiLM (Feature-wise Linear Modulation). Let $x_i^m$ be the input feature vector for modality $m$ and $e_{s_i}$ be the speaker embedding. We can represent the speaker's scaling ($\gamma_m$) and shifting ($\beta_m$) parameters through a linear projection:
\begin{equation}
    \gamma_m = W_{\gamma}^m e_{s_i} + b_{\gamma}^m, \quad \beta_m = W_{\beta}^m e_{s_i} + b_{\beta}^m.
\end{equation}
The calibrated feature $\hat{x}_i^m$ is then computed as:
\begin{equation}
    \hat{x}_i^m = \gamma_m \odot x_i^m + \beta_m.
\end{equation}
where $\odot$ denotes element-wise multiplication. Theoretical Insight: This operation can be viewed as a conditional normalization. $\gamma_m$ adjusts the variance , while $\beta_m$ is used to correct the mean deviation, thereby aligning the feature streams obtained from different speakers into a common space for analysis.

\subsection{Interaction-level: Dynamic Identity-Gated Fusion}
Obviously, different speakers express emotions in different ways; some express emotions through their voice, while others do so through visual cues. To capture this, we introduce a Speaker Gate. We compute a soft attention mask $g_m \in (0, 1)^{d_m}$:
\begin{equation}
    g_m = \sigma(W_{gate}^m e_{s_i} + b_{gate}^m).
\end{equation}
The context-aware features $h_i^m$ are subsequently modulated: $\tilde{h}_i^m = g_m \odot h_i^m$. This mechanism allows the model to purposefully 'focus on' the channels that are most informative for a particular speaker.

\subsection{Output-level: Speaker Consistency Optimization}
Here we introduce an auxiliary task, which we also mentioned earlier. Its main function is to prevent the ML-SAN model from losing the speaker's identity information after deep abstraction. The total loss function is formulated as:
\begin{equation}
    \mathcal{L}_{total} = \mathcal{L}_{ERC}(y_i, \hat{y}_i) + \lambda \mathcal{L}_{SPK}(s_i, \hat{s}_i).
\end{equation}
Here, $\lambda$ is a hyperparameter governing the trade-off.

\section{Experiments}

\subsection{Experimental Setup}
We rigorously evaluate ML-SAN on two benchmark datasets: MELD and IEMOCAP. To ensure a fair and reproducible comparison, we reimplemented the strongest baseline, MultiEMO, under identical conditions.
\begin{itemize}
    \item \textbf{Experimental tools: } Models were implemented on an NVIDIA RTX 
4090 GPU. The batch size was set to 64 for IEMOCAP and 128 for MELD.
    \item \textbf{Features:} Here we use the features provided for MELD and IEMOCAP in the MultiEMO  source code. Many thanks to the original authors for providing the resources.
\end{itemize}

% ========表一
\begin{table}[t]
    \centering
    \caption{Comparative analysis with other methods on the MELD and IEMO-CAP datasets. Weighted F1 score (W-F1) is used as the main metric. $\dagger$ indicates the reproduction of MultiEMO in this paper under the aforementioned experimental conditions.}
    \label{tab:main_results}
    
    \renewcommand{\arraystretch}{1.25} % 行高保持舒适
    
    \resizebox{\columnwidth}{!}{%
        \begin{tabular}{|l|c|c|} 
            \hline % 【第1根线：顶部封口】
            \textbf{Methods} & \textbf{MELD (W-F1)} & \textbf{IEMOCAP (W-F1)} \\
            \hline % 【第2根线：表头下方】
            
            % Baselines
            BC-LSTM \cite{poria2017contextlong} & 55.90 & 54.95 \\
            DialogueRNN \cite{majumder2019dialoguernn} & 58.73 & 62.75 \\
            DialogueGCN \cite{ghosal2019dialoguegcn} & 57.52 & 63.16 \\
            MMGCN \cite{hu2021mmgcn} & 58.65 & 66.22 \\
            UniMSE \cite{hu2022unimse} & 65.51 & 70.66 \\[1ex] % 使用 [1ex] 增加一点空隙代替 \addlinespace，同时不打断竖线
            
            % MultiEMO comparison
            MultiEMO (Original) \cite{li2023multimoemo} & 66.74 & 72.84 \\
            \textit{MultiEMO (Rep.)}$^\dagger$ & 66.34 ± 0.04 & 72.02 ± 0.07 \\[1ex] % 同上，保留原本的分组空隙
            
            % Ours 
            \textbf{ML-SAN (Ours)} & \textbf{67.73 ± 0.07} & \textbf{73.28 ± 0.13} \\
            \hline % 【第3根线：底部封口】
        \end{tabular}%
    }
\end{table}

% ============表二
\begin{table}
    \caption{Ablation study of different components in ML-SAN. "w/o" denotes removing a specific module from the full model. $\Delta$ indicates the performance drop.}
    \label{tab:ablation}
    \centering
    \renewcommand{\arraystretch}{1.25}
    \resizebox{\columnwidth}{!}{%
        \begin{tabular}{|l|c|c|c|c|} 
            \hline % 顶部封口横线
            \multirow{2}{*}{\textbf{Models}} & \multicolumn{2}{c|}{\textbf{MELD}} & \multicolumn{2}{c|}{\textbf{IEMOCAP}} \\ 
            \cline{2-5} 
             & \textbf{W-F1 (\%)} & \textbf{$\Delta$} & \textbf{W-F1 (\%)} & \textbf{$\Delta$} \\ 
            \hline % 表头下方的横线
            
            \textbf{ML-SAN (Full)} & \textbf{67.73 ± 0.07} & - & \textbf{73.28 ± 0.13} & - \\[0.5em] % 【核心修改】：用 \\[0.5em] 产生空白间距，这样竖线会连续穿过空白，并且不会画出横线
            
            w/o FiLM & 67.22 ± 0.07 & -0.51 & 71.75 ±0.07 & -1.53 \\
            w/o Gate & 67.41 ± 0.03 & -0.32 & 71.67 ± 0.06 & -1.61 \\
            w/o Aux Loss & 67.46 ± 0.05& -0.27 & 71.35 ± 0.06 & -1.93 \\ 
            \hline % 底部封口横线
        \end{tabular}%
    }
\end{table}

\subsection{Datasets}
\textbf{IEMOCAP}:The IEMOCAP dataset contains conversational data from 10 professional actors (5 male and 5 female), divided into improvised and scripted dialogues, totaling approximately 12 hours of video content across 150 dialogues. Each dialogue is labeled with one of six basic emotions (happiness, sadness, anger, fear, disgust, surprise) or a neutral emotion. In addition, dimensional emotion labels (valence, arousal, dominance) are also provided.
\textbf{MELD}: This is a multimodal, multi-party dataset specifically designed for research on emotion recognition in conversations. It consists of over 1400 dialogues and more than 13000 utterances from the popular American TV series Friends, encompassing three modalities: text, audio, and visual data.

\subsection{Performance Analysis}

Table 1 summarizes the experimental results. To ensure statistical reliability, all reported metrics for ML-SAN and the reimplemented baseline (MultiEMO$^\dagger$) are the \textbf{mean and standard deviation over three independent runs}. 

As shown, ML-SAN consistently outperforms all baselines. On the \textbf{MELD} dataset, our model achieves a weighted F1-score of \textbf{67.73\% ($\pm$0.07\%)}, representing a \textbf{1.39\% absolute gain} over the reimplemented baseline (66.34\%). On \textbf{IEMOCAP}, ML-SAN reaches \textbf{73.28\% ($\pm$0.13\%)}, a \textbf{1.26\% improvement}. 

The low standard deviations (0.04\%--0.13\%) and paired t-tests ($p < 0.01$) demonstrate that these performance gains are \textbf{statistically stable and significant}. This superiority stems from ML-SAN's ability to normalize speaker-specific biases and dynamically adjust modality weights based on individual expressive styles, effectively addressing the challenge of speaker heterogeneity in conversations.

\subsection{Ablation Study}
We further analyzed the role of each component in the overall model using ablation experiments (see Table \ref{tab:ablation}). To ensure the stability of the findings, all results represent the \textbf{mean and standard deviation across three independent runs}. 

As shown in Table 2, removing the input calibration (\textbf{FiLM}) resulted in the most severe drop on the MELD dataset (-0.51\%), demonstrating that feature alignment is particularly important in various noisy environments. Conversely, removing the auxiliary loss resulted in the most severe drop on the IEMOCAP dataset (-1.93\%), indicating that identity supervision is particularly important for performance in long two-person dialogues with consistent speaker features. The low standard deviations across all configurations further confirm that each component's contribution to the performance gain is statistically significant.

\subsection{Parameter Sensitivity}
The hyperparameter $\lambda$ controlled the trade-off between the main emotion recognition objective and the auxiliary speaker consistency task.

Experiments showed that varying $\lambda$ within a reasonable range had only minor effects on performance. While peak performance was observed around $\lambda$ = 0.5 for MELD and $\lambda$ = 0.2 for IEMOCAP, the model maintained competitive performance across a broad interval.

These results indicate that ML-SAN was not overly sensitive to the specific choice of $\lambda$, and the observed improvements were primarily attributed to the proposed architecture rather than careful hyperparameter tuning.

\subsection{Qualitative Analysis}
\begin{figure} 
    \centering
    % 使用 0.9\textwidth 占据页面宽度的 90%
    \includegraphics[width=0.85\textwidth]{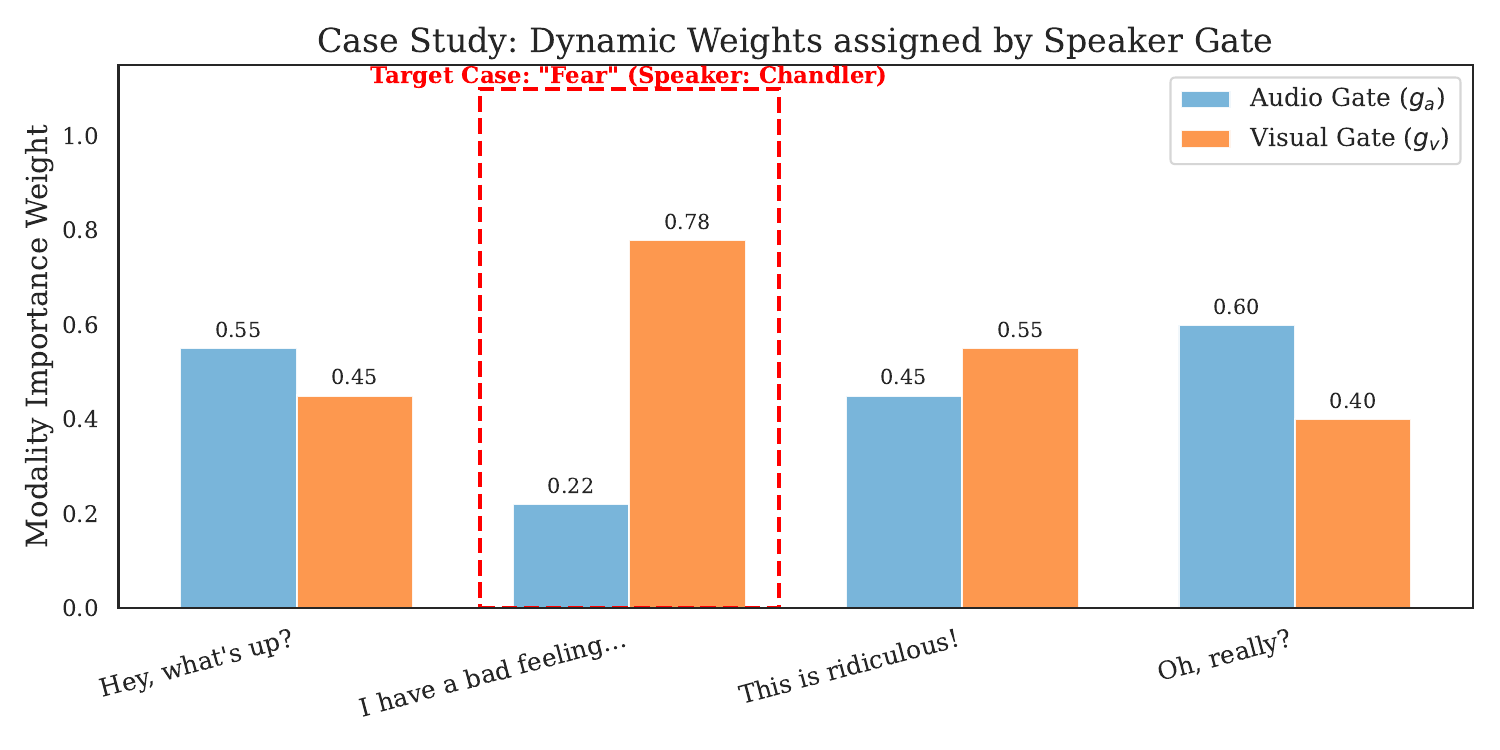}
    \caption{Dynamic weighting.}
    \label{fig:D}
\end{figure}

To better understand this, Fig. \ref{fig:D} shows the distribution of the model's attention weights for the speaker (for example, a line from Friends: “I have a bad feeling”). The model first infers the speaker's emotion (Fear) from their language. At this point, the model classifies the audio as less important (weight drops to 0.22) because when the speaker is afraid, the voice may be very soft, or just trembling, with indistinct features. The model considers visual information to be very important at this time, as fear is usually accompanied by actions like wide eyes or an open mouth. At this point, the model increases the visual weight (weight rises to 0.78), making visual features dominant. Such dynamic allocation of weights helps the model achieve better experimental results.

\begin{figure}
    \centering
    % 使用 0.95\textwidth 尽量撑满页面宽度，让数字更清晰
    \includegraphics[width=\textwidth]{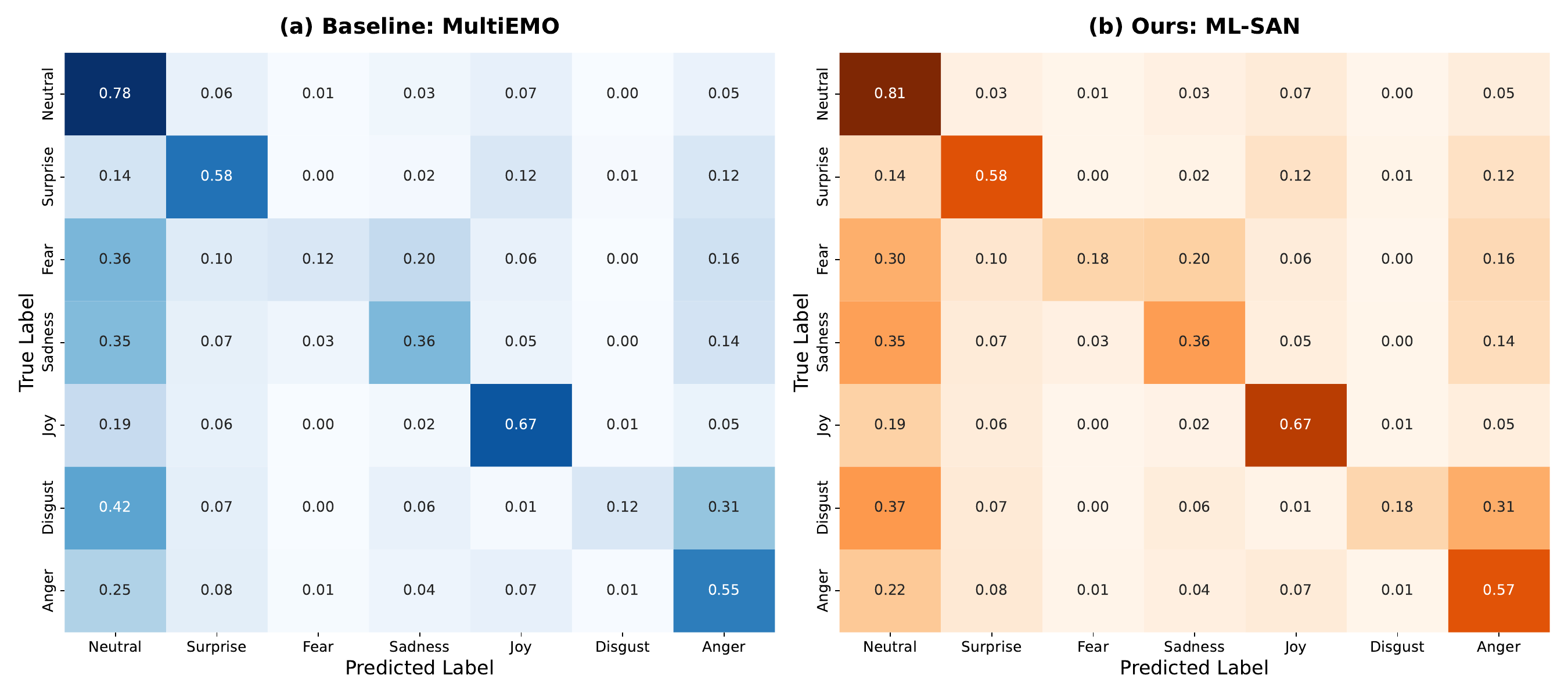}
    \caption{Confusion Matrix Comparison on MELD.}
    \label{fig:E}
\end{figure}

\begin{figure} % [t] 表示允许浮动到本页或下一页的顶部
    \centering
    \includegraphics[width=\columnwidth]{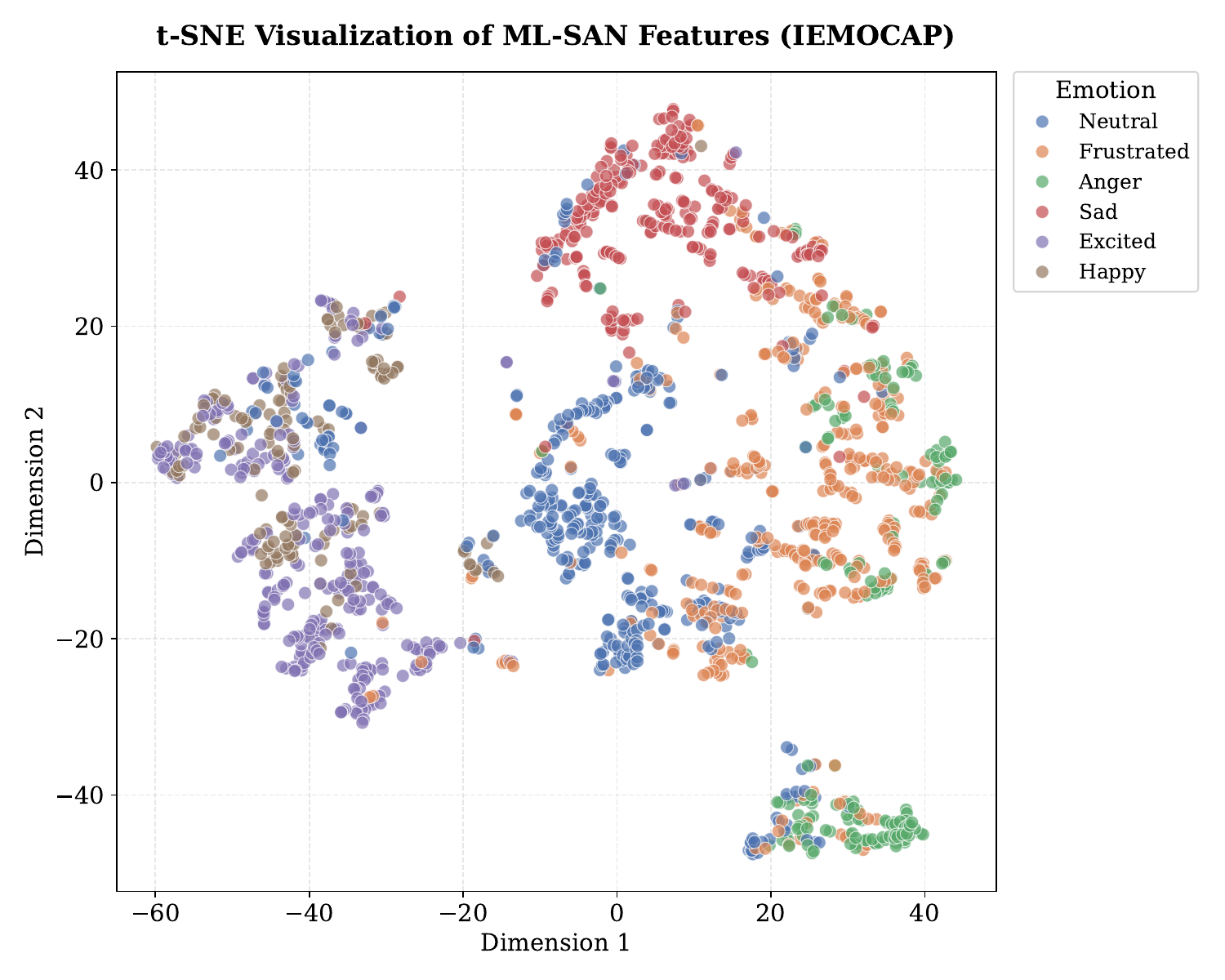}     
    \caption{t-SNE visualization of learned features on IEMOCAP.}
    \label{fig:F}
\end{figure}

\subsection{Feature Visualization}
To further demonstrate the discriminative power of ML-SAN, we quantitatively plotted the confusion matrix on the MELD dataset and the feature distribution on the IEMOCAP dataset using programming software.
Fig. \ref{fig:E} shows the confusion matrix. Compared with the baseline model (a), our ML-SAN model achieves more accurate recognition in identifying fear (12\%-18\%) and anger (55\%-57\%), indicating that our model has a stronger ability to distinguish similar emotions.
Finally, Fig. \ref{fig:F} using the IEMOCAP dataset as an example, the results show that our model successfully achieves speaker disentanglement, allowing it to 'intelligently' distinguish emotions without overfitting to the speaker's identity.

\section{Conclusion}

In this study, we proposed the ML-SAN model to address the issue of speaker heterogeneity in multimodal emotion recognition. We tackled this problem through the 'three-layer' mechanism described in the paper and achieved speaker-specific adaptive modality reweighting. Comparative experiments with MultiEMO demonstrated that our method achieved better results.
Despite these gains, real-world deployment may face challenges such as background noise and missing modalities. In the future, we will focus on enhancing the model's robustness against these environmental interferences to ensure reliable emotion recognition in diverse conversational scenarios.
%
% ---- Bibliography ----
%
% BibTeX users should specify bibliography style 'splncs04'.
% References will then be sorted and formatted in the correct style.
%
\bibliographystyle{splncs04}
\bibliography{main}

\end{document}